\documentclass
[12pt]{article}
\usepackage{amssymb}
\usepackage{epsfig}

\catcode`\@=11
\def\numberbysection{\@addtoreset{equation}{section}
        \def\theequation{\thesection.\arabic{equation}}}

\def\beq{\begin{equation}}
\def\eeq{\end{equation}}
\numberbysection
\begin{document}
\begin{titlepage}
\begin{center}
\hfill  \\
\vskip 1.in {\Large \bf AdS spacetime in Lorentz covariant gauges} \vskip 0.5in P. Valtancoli
\\[.2in]
{\em Dipartimento di Fisica, Polo Scientifico Universit\'a di Firenze \\
and INFN, Sezione di Firenze (Italy)\\
Via G. Sansone 1, 50019 Sesto Fiorentino, Italy}
\end{center}
\vskip .5in
\begin{abstract}
We show how to generate the AdS spacetime metric in general Lorentz covariant gauges. In particular we propose an iterative method for solving the Lorentz gauge.
\end{abstract}
\medskip
\end{titlepage}
\pagenumbering{arabic}
\section{Introduction}

Recently there has been a renewed interest in the cosmological constant of general relativity, since it could be useful as a cheap explanation for the antigravity force which drives the accelerated expansion of the universe, better known as dark energy \cite{1}-\cite{2}. It is now widely accepted that the cosmological constant is a very small parameter, analogously to what happens for the neutrino mass, and related to the vacuum energy of the quantum fields \cite{3}.

The $AdS$ space-time \cite{4} is usually represented in a Lorentz non-covariant form, but there are physical applications like for example the study of cosmological gravitational waves \cite{5}, in which it is useful to reformulate it in a Lorentz covariant form. It is the purpose of this paper to generate all the Lorentz covariant solutions for $AdS$ space-time. We start from solving perturbatively the Einstein equations up to the second order in $\Lambda$ in the Lorentz gauge, which turns out to be a tedious calculation.

To simplify the discussion we are going to present a general $5$-dimensional representation of the $AdS$ space-time in terms of an arbitrary function $ f( \Lambda x^2 )$, where $ x^2 \ = \ \eta^{\mu\nu} x_\mu x_\nu $. We then show that the Lorentz gauge fixes this arbitrary function with a recursive method and the result agrees with the direct perturbative solution of the Einstein equations.

\section{Perturbative calculation of the Einstein equations}

We are going to discuss all the aspects of the AdS space-time in the Lorentz gauge. We start performing the perturbative calculation directly using the Einstein equations, and only afterwards we will introduce a faster method. We recall the known perturbative solution of the $AdS$ space-time at the first order in $\Lambda $

\begin{eqnarray} g_{\mu\nu} & = & \eta_{\mu\nu} \ + \ h_{\mu\nu}^{(1) \Lambda} \ + O(\Lambda^2)\  \nonumber \\
h_{\mu\nu}^{(1) \Lambda} & = & - \frac{\Lambda}{9} \ ( \ x_\mu x_\nu  +  2  \eta_{\mu\nu}  x^2 \ )
\label{21} \end{eqnarray}

It is straightward to show that eq. (\ref{21}) satisfies the Lorentz gauge:

\beq  \eta^{\mu\mu'} \partial_{\mu'} h_{\mu\nu}^{(1) \Lambda} \ = \ \frac{1}{2} \partial_\nu ( \eta^{\mu\mu'} h_{\mu\mu'}^{(1) \Lambda} ) \label{22}
\eeq

The associated connection is, at this order, given by:

\beq \Gamma_{\alpha\beta}^{ \mu (1) \Lambda} \ = \ - \frac{\Lambda}{9} ( \ 2 \delta^\mu_\alpha x_\beta + 2 \delta^\mu_\beta x_\alpha - \eta_{\alpha\beta} x^\mu \ ) \
\label{23} \eeq

We can easily compute the curvature tensor, limited at the linear term in the connection

\beq  R_{\mu\beta\nu}^{ \alpha (1) \Lambda} \ = \  \frac{\Lambda}{3} ( \ \eta_{\beta\nu} \delta_\mu^\alpha \ - \ \eta_{\beta\mu} \delta_\nu^\alpha \  ) \label{24} \eeq

The corresponding Ricci tensor and the curvature are then

\beq R_{\beta\nu}^{(1) \Lambda} \ = \ \Lambda \ \eta_{\beta\nu} \ \ \ \   R^{(1) \Lambda} \ = \ \eta^{\beta\nu} R_{\beta\nu}^{(1) \Lambda} \ = \ 4 \Lambda \label{25} \eeq

solving perturbatively the Einstein equations.

Now we are going to compute the perturbative metric  at the second order in $ \Lambda $

\beq g_{\mu\nu} \ = \ \eta_{\mu\nu} \ + \ h_{\mu\nu}^{(1) \Lambda} \ + \ h_{\mu\nu}^{(2) \Lambda} \ + \ O(\Lambda^3)
\label{26} \eeq

We can again impose the Lorentz gauge

\beq \eta^{\mu\mu'} \partial_{\mu'} h_{\mu\nu}^{(2) \Lambda} \ = \ \frac{1}{2} \partial_\nu ( \eta^{\mu\mu'} h_{\mu\mu'}^{(2) \Lambda} ) \label{27} \eeq

We must be careful that the connection is made by several pieces

\begin{eqnarray} \Gamma_{\alpha\beta}^{ \mu (2) \Lambda} & = & \Gamma_{\alpha\beta}^{ \mu (2) \Lambda \ I} \ + \ \Gamma_{\alpha\beta}^{ \mu (2) \Lambda \ II} \nonumber \\
\Gamma_{\alpha\beta}^{ \mu (2) \Lambda \ I} & = & \frac{1}{2} \ \eta^{\mu\mu'} \ ( \ \partial_\alpha h_{\mu'\beta}^{(2) \Lambda} \ + \ \partial_\beta h_{\mu'\alpha}^{(2) \Lambda} \ - \
\partial_{\mu'} h_{\alpha\beta}^{(2) \Lambda} \ ) \nonumber \\
\Gamma_{\alpha\beta}^{ \mu (2) \Lambda \ II} & = & - \frac{1}{2} \ h^{\mu \mu'}_{(1) \Lambda} \ ( \ \partial_\alpha h_{\mu'\beta}^{(1) \Lambda} \ + \ \partial_\beta h_{\mu'\alpha}^{(1) \Lambda} \ - \
\partial_{\mu'} h_{\alpha\beta}^{(1) \Lambda} \ ) \label{28} \end{eqnarray}

where $ h^{\mu \nu}_{(1) \Lambda} \ = \ \eta^{\mu\mu'} \eta^{\nu\nu'} \ h_{\mu'\nu'}^{(1) \Lambda} $.

Let us work out the second term

 \beq \Gamma_{\beta\nu}^{ \alpha (2) \Lambda \ II} \ = \ - \frac{\Lambda^2}{81} \ ( \ 4 x^\alpha x_\beta x_\nu - 3 \eta_{\beta\nu} x^\alpha x^2 + 4 \delta^\alpha_\beta x_\nu x^2 + 4 \delta^\alpha_\nu x_\beta x^2 \ )
 \label{29} \eeq

Its contribution to the curvature tensor is given by

\begin{eqnarray} R_{\beta\mu\nu}^{ \alpha (2) \Lambda \ II} & = & \partial_\mu \Gamma_{\beta\nu}^{ \alpha (2) \Lambda \ II} \ - \ ( \mu \leftrightarrow \nu ) \ = \ \nonumber \\
& = & - \frac{\Lambda^2}{81} [ \ 10 \eta_{\beta\mu} x_\nu x^\alpha - 10 \eta_{\beta\nu} x_\mu x^\alpha + 4 \delta^\alpha_\nu x_\beta x_\mu \nonumber \\
& \ & - 4 \delta^\alpha_\mu x_\beta x_\nu + 7 \delta^\alpha_\nu \eta_{\beta\mu} x^2 - 7 \delta^\alpha_\mu \eta_{\beta\nu} x^2
\ ] \label{210} \end{eqnarray}

The Ricci tensor

 \beq R_{\beta\nu}^{ (2) \Lambda \ II} \ = \ \delta^\mu_\alpha \ R_{\beta\mu\nu}^{ \alpha (2) \Lambda \ II} \ = \ \frac{\Lambda^2}{81} ( \  2 x_\beta x_\nu + 31 \eta_{\beta\nu} x^2 \ )
  \label{211} \eeq

and its contribution to the Einstein equation is

\beq R_{\beta\nu}^{ (2) \Lambda \ II} \ - \ \frac{1}{2} R^{ (2) \Lambda \ II} \eta_{\beta\nu} \ = \ \frac{2 \Lambda^2}{81} ( \ x_\beta x_\nu - 16 \eta_{\beta\nu} x^2 \ )
\label{212} \eeq

The first term of the connection produces the following curvature tensor $ \Gamma_{\alpha\beta}^{ \mu (2) \Lambda \ I} $

\beq R_{\beta\nu}^{ (2) \Lambda \ I} \ = \ - \ \frac{1}{2} \eta^{\rho\rho'} \partial_\rho \partial_{\rho'} h_{\beta\nu}^{ (2) \Lambda} \ + \ { \rm gauge \  terms  }
\label{213} \eeq

giving rise to the another contribution to the Einstein equations

\beq R_{\beta\nu}^{ (2) \Lambda \ I} \ - \ \frac{1}{2} R^{ (2) \Lambda \ I} \eta_{\beta\nu} \ = \ - \ \frac{1}{2} \eta^{\rho\rho'} \partial_\rho \partial_{\rho'} \ \left( \ h_{\beta\nu}^{ (2) \Lambda} - \frac{1}{2} \eta_{\beta\nu} h^{ (2) \Lambda} \right)
 \label{214} \eeq

Till now we have worked out in the curvature tensor only linear terms in the connections. Let us add now the non linear ones:

\begin{eqnarray}
R_{\beta\mu\nu}^{ \alpha (2) \Lambda \ III} & = & \Gamma_{\mu\alpha'}^{ \alpha (1) \Lambda} \ \Gamma_{\beta\nu}^{ \alpha' (1) \Lambda} \ - \ ( \mu \leftrightarrow \nu ) \ = \ \nonumber \\
& = & \frac{\Lambda^2}{81} [ 4 \delta^\alpha_\mu x_\nu x_\beta - 2 \delta^\alpha_\mu \eta_{\beta\nu} x^2 - 4 \delta^\alpha_\nu x_\mu x_\beta + \nonumber \\
& \ & + 2 \delta^\alpha_\nu \eta_{\beta\mu} x^2 - \eta_{\beta\mu} x^\alpha x_\nu + \eta_{\beta\nu} x^\alpha x_\mu ]
\label{215} \end{eqnarray}

The associated Ricci tensor is

\beq R_{\beta\nu}^{(2) \Lambda \ III} \ = \ \frac{\Lambda^2}{81}( \ 11 x_\nu x_\beta - 5 \eta_{\beta\nu} x^2 \ )
\label{216} \eeq

giving rise to the third contibution to the Einstein equations

\beq   R_{\beta\nu}^{ (2) \Lambda \ III} \ - \ \frac{1}{2} R^{ (2) \Lambda \ III} \eta_{\beta\nu} \ = \ \frac{\Lambda^2}{81}( \ 11 x_\nu x_\beta - \frac{1}{2} \eta_{\beta\nu} x^2 \ )
\label{217} \eeq

We have not yet finished. We must add some residual extra contributions, given by

\beq \frac{1}{2} R^{ (2) \Lambda \ IV} \ = \ h^{(1) \Lambda \beta \nu} R_{\beta\nu}^{(1) \Lambda} \ \ \ \ \ \rightarrow \ \ \ \ \ - \frac{1}{2} R^{ (2) \Lambda \ IV} \eta_{\beta\nu} \ = \
- \frac{\Lambda^2 x^2}{2} \eta_{\beta\nu}
\label{218} \eeq

and finally

\beq  - \frac{1}{2} R^{(1)\Lambda} h^{(1) \Lambda}_{\beta \nu} + \Lambda h^{(1) \Lambda}_{\beta \nu} \ = \ - \Lambda h^{(1) \Lambda}_{\beta \nu}
\label{219} \eeq

By collecting all the various ( five ) terms

\begin{eqnarray} & - & \frac{1}{2} \eta^{\rho\rho'} \partial_\rho \partial_{\rho'} \ \left( \ h_{\beta\nu}^{ (2) \Lambda} - \frac{1}{2} \eta_{\beta\nu} h^{ (2) \Lambda} \right) \ + \
\frac{2 \Lambda^2}{81} ( \ x_\beta x_\nu - 16 \eta_{\beta\nu} x^2 \ ) \nonumber \\
& + & \frac{\Lambda^2}{81}( \ 11 x_\nu x_\beta - \frac{1}{2} \eta_{\beta\nu} x^2 \ ) \ - \ \frac{\Lambda^2 x^2}{2} \eta_{\beta\nu} \ + \ \frac{\Lambda^2}{9} ( x_\beta x_\nu + 2 \eta_{\beta\nu} x^2 ) \ = \ 0
\label{220}
\end{eqnarray}

we arrive at the final equation

\beq
 \eta^{\rho\rho'} \partial_\rho \partial_{\rho'} \ \left( \ h_{\beta\nu}^{ (2) \Lambda} - \frac{1}{2} \eta_{\beta\nu} h^{ (2) \Lambda} \right) \ = \ \frac{22}{81} ( 2 x_\beta x_\nu - 5 \eta_{\beta\nu} x^2 )
  \label{221} \eeq

The compatibility between this equation and the Lorentz gauge is straightforward since applying the operator $\partial_\beta \ ( \ {\rm  or } \ \partial_\nu )$ to both sides of this equation we simply get zero.

Finally we obtain the solution of this equation at the second order in $\Lambda$:

\begin{eqnarray}
h^{(2)\Lambda}_{\beta\nu} \ = \ \frac{11}{16 \cdot 81} \Lambda^2 (  \  4 x_\beta x_\nu x^2 + 5 \eta_{\beta\nu} x^4 ) \nonumber \\
h^{(2)\Lambda} \ = \ \eta^{\beta\nu} h^{(2)\Lambda}_{\beta\nu} \ = \ \frac{11}{54} \Lambda^2 x^4 \ \ \ \ \ \ x^4 \equiv (x^2)^2
\label{222}
\end{eqnarray}

We have just learned that the Lorentz gauge imposes strong constraints on the general solution. Let us suppose to make the perturbative calculation at the generic order $n$:

\beq
g_{\mu\nu} \ = \ \eta_{\mu\nu} \ + \ h^{(1)\Lambda}_{\mu\nu} \ + \ .... \ + \ h^{(n)\Lambda}_{\mu\nu} \ + \ ....
\label{223}
\eeq

the final equation, similar to eq. ( \ref{221} ), at the order $n$ is of the type

\beq
 \eta^{\rho\rho'} \partial_\rho \partial_{\rho'} \ \left( \ h_{\beta\nu}^{ (n) \Lambda} - \frac{1}{2} \eta_{\beta\nu} h^{ (n) \Lambda} \right) \ = \  ( \ A_n x_\beta x_\nu x^{2(n-2)} +
  B_n \eta_{\beta\nu} x^{2(n-1)} \ )
\label{224}
\eeq

Let us apply the operator $\partial_\beta \ ( \ {\rm  or } \ \partial_\nu )$ to both sides of this equation

\beq
\partial_\beta ( ... ) \ = \ 0 \ \ \ \ \ \ \rightarrow \ \ \ \ \ \ \ B_n \ = \ - \frac{2n + 1}{2(n-1)} \ A_n
\label{225}
\eeq

the Lorentz gauge implies the constraint

\beq
\eta^{\rho\rho'} \partial_\rho \partial_{\rho'} \ \left( \ h_{\beta\nu}^{ (n) \Lambda} - \frac{1}{2} \eta_{\beta\nu} h^{ (n) \Lambda} \right) \ = \  c_n \Lambda^n [ \
2(n-1) x_\beta x_\nu x^{2(n-2)} - ( 2n+1 ) \eta_{\beta\nu} x^{2(n-1)} \ ]
\label{226} \eeq

Therefore the solution at the order $n$ is of the type

\begin{eqnarray}
h^{(n)\Lambda}_{\beta\nu} \ = \ c_n \Lambda^n \left[  \  \frac{ x_\beta x_\nu x^{2(n-1)} }{2(n+2)} \ + \ \frac{ ( n+3 ) \eta_{\beta\nu} x^{2n} }{4n(n+2)} \right] \nonumber \\
h^{(n)\Lambda} \ = \ \eta^{\beta\nu} h^{(n)\Lambda}_{\beta\nu} \ = \ \frac{3}{2n} c_n \Lambda^n x^{2n} \ \ \ \ \ \ x^{2n} \equiv (x^2)^n
\label{227}
\end{eqnarray}

Unfortunately the $c_n$ coefficients cannot be computed only from the Lorentz gauge and must be determined by solving the Einstein equation.

\section{General solution of the Einstein equations}

To solve the Einstein equation, we are going to embed the $AdS$ space-time in the following $5d$ space

\beq ds^2 \ = \ d X_0^2 \ + \  d X_1^2 \ - \  d X_2^2 \ - \ d X_3^2 \ - \ d X_4^2
\label{31}\eeq

subject to the constraint

\beq
X^2_0 \ + \ \eta^{ij} X_i X_j \ = \ \frac{3}{\Lambda} \ \ \ \ \ \ i = 1, ..., 4
\label{32}\eeq

We can automatically generate solutions of the Einstein equations simply determining the mapping $ X_0 = X_0 ( x_\mu ), X_i = X_i ( x_\mu ) $. In this choice we must respect the
Lorentz covariance of the $4d$ space-time and we are forced to choose this mapping

\begin{eqnarray}
X_0 & = & \sqrt{ \frac{3}{\Lambda} } \ \sqrt{ 1 \ - \frac{\Lambda}{3} x^2 f^2 ( \Lambda x^2 ) } \nonumber \\
X_i & = & x_i \ f ( \Lambda x^2 )
\label{33} \end{eqnarray}

with the $4d$ metric defined as usual by the formula

\beq
g_{\mu\nu} \ = \ \eta^{a b} \ \frac{d X_a}{ d x^\mu} \ \frac{d X_b}{ d x^\nu}
\label{34} \eeq

All these mappings produce solutions of the Einstein equation in a covariant gauge and depend on an arbitrary function $f ( \Lambda x^2 )$. First let us study the basic solution with
$f ( \Lambda x^2 ) = 1 $:

\begin{eqnarray}
X_0 & = &  \sqrt{ \frac{3}{\Lambda} } \ \sqrt{ 1 \ - \frac{\Lambda}{3} x^2  } \ \ \ \ \ \ X_i \ = \ x_i \nonumber \\
g^{(0)}_{\mu\nu}  & = & \eta_{\mu\nu} \ + \ \frac{\frac{\Lambda}{3} x_\mu x_\nu}{ 1 - \frac{\Lambda}{3} x^2 }
\label{35} \end{eqnarray}
We can easily verify that it solves the Einstein equations exactly. The inverse metric is given by

\beq g^{\mu\nu (0)} \ = \ \eta^{\mu\nu} \ - \ \frac{\Lambda}{3} x^\mu x^\nu
\label{36} \eeq

We can compute the associated connection

\beq
\Gamma^{\mu (0)}_{\alpha\beta} \ = \ \frac{\Lambda}{3} x^\mu g^{(0)}_{\alpha\beta}
\label{37} \eeq

and the complete curvature tensor

\begin{eqnarray}
R^{\alpha (0)}_{\beta\mu\nu} & = &  \partial_\mu \Gamma^{\alpha (0)}_{\beta\nu} \ + \ \Gamma^{\alpha (0)}_{\sigma\mu} \ \Gamma^{\sigma (0)}_{\beta\nu} \ - \ ( \mu \leftrightarrow \nu ) \ = \ \nonumber \\
& = & \frac{\Lambda}{3} ( \ \delta^\alpha_\mu g^{(0)}_{\beta\nu} \ - \ \delta^\alpha_\nu g^{(0)}_{\beta\mu} \ )
\label{38} \end{eqnarray}

Analogously the Ricci tensor and the curvature have simple forms

\beq R^{(0)}_{\beta\nu} \ = \ \Lambda \ g^{(0)}_{\beta\nu}  \ \ \ \ \ R^{(0)} \ = \ 4 \Lambda
\label{39} \eeq

solving exactly the Einstein equations. Let us note that this solution ( similarly to those with an arbitrary function $ f(\Lambda x^2 )$ ) is singular for large events of the order

\beq x^2 \simeq \frac{1}{\Lambda}
\label{310} \eeq

This singularity is present both for positive and negative cosmological constant.

It is possibile fixing the arbitrary function $ f(\Lambda x^2 )$ adding into this scheme the Lorentz gauge condition

\beq \eta^{\mu\mu'} \ \partial_{\mu'} h_{\mu\nu} \ = \ \frac{1}{2} \ \partial_\nu ( \ \eta^{\mu\mu'} h_{\mu\mu'} \ )
\label{311} \eeq

By substituting the definition of the metric in terms of the mapping $ X^a = X^a  ( x_i ) $ we get

\beq
\eta^{ a b } \eta^{\mu\mu'} \ \partial_{\mu}\partial_{\mu'} X_a \cdot \partial_\nu X_b \ = \ 0
\label{312}
\eeq

The variables $X_a$ are constrained leading to the property $ \eta^{ a b } X_a \cdot \partial_\nu X_b \ = \ 0 $,  therefore we must impose that

\beq
\eta^{\mu\mu'} \ \partial_{\mu}\partial_{\mu'} X_a \ = \ \lambda ( \Lambda x^2 ) \ X_a
 \label{313} \eeq

where we have introduced a second function $ \lambda ( \Lambda x^2 ) $. We are going to show that this equation (\ref{313}) determines univocally both functions $ f(\Lambda x^2 )$ and
$ \lambda ( \Lambda x^2 ) $.

Let us introduce the following power series

\begin{eqnarray}
\lambda ( \Lambda x^2 ) & = & \sum_{n=1}^\infty \lambda_n \Lambda^n x^{2(n-1)} \nonumber \\
f ( \Lambda x^2 ) & = & 1 \ + \ \sum_{n=1}^\infty f_n \Lambda^n x^{2n}
\label{314} \end{eqnarray}

The coefficients $\lambda_n$ are determined by the eq. for $X_0$ while the coefficients $f_n$ are determined by the eq. for $X_i$ in a recursive scheme. We start from the lowest order

\beq
X_0 \simeq \sqrt{\frac{3}{\Lambda}} \ \left( \  1 - \frac{\Lambda}{6} x^2  \right) \ \ \ \ \ \rightarrow \ \ \ \ \ \lambda_1 \ = \ - \ \frac{4}{3}
\label{315} \eeq

The coefficient $\lambda_1$ allows to compute the coefficient $f_1$ by using the equation for $X_i$

\beq
\eta^{\rho\rho'} \ \partial_{\rho}\partial_{\rho'} X_i \simeq \ \eta^{\rho\rho'} \ \partial_{\rho}\partial_{\rho'} ( x_i \cdot f_1 \Lambda x^2 ) \ = \ - \ \frac{4}{3} \Lambda x_i
\label{316} \eeq

from which we obtain

\beq
f_1 \ = \ - \ \frac{1}{9} \ \ \ \ \ \rightarrow \ \ \ \ f(\Lambda x^2 ) \ = \ 1 - \frac{\Lambda}{9} x^2 \ + \ O(\Lambda^2 x^4)
\label{317} \eeq

Going back to the eq. for $X_0$ at the next order in $\Lambda$, we can use $f_1$ to determine $\lambda_2$ as follows

\beq
X_0 \simeq \sqrt{\frac{3}{\Lambda}} \ \left( \  1 - \frac{\Lambda}{6} x^2  +  \frac{5}{9 \cdot 24} \Lambda^2 x^4 + O(\Lambda^3 x^6) \right)
\label{318} \eeq

from which we compute $\lambda_2 = \frac{1}{3}$ and therefore

\beq \lambda ( \Lambda x^2 ) \ = \ - \  \frac{4}{3} \Lambda + \frac{1}{3} \Lambda^2 x^2 + O ( \Lambda^3 x^4 )
\label{319} \eeq

Repeating the iteration, the coefficient $\lambda_2$ determines $f_2$ at the next order in $\Lambda$ by using the equation for $X_i$

\beq f_2 \ = \ \frac{13}{27 \cdot 32 }
\label{320} \eeq

from which we know $X_i$ up to the second order in $\Lambda$

\beq
X_i \simeq x_i \left( 1 - \frac{\Lambda}{9} x^2 +  \frac{13}{27 \cdot 32} \Lambda^2 x^4 + O(\Lambda^3 x^6) \right)
\label{321} \eeq

By substituting these findings in the equation (\ref{34}), defining the metric in terms of the mapping $X_a(x_i)$

\beq
g_{\mu\nu} \ = \ \eta^{a b} \ \frac{d X_a}{ d x^\mu} \ \frac{d X_b}{ d x^\nu} \ = \ \eta_{\mu\nu} \ + \ h_{\mu\nu}^{(1)\Lambda} \ + \ h_{\mu\nu}^{(2)\Lambda} + ....
\label{322} \eeq

we recover the perturbative solutions (\ref{21}) e (\ref{222})

\begin{eqnarray}
h_{\mu\nu}^{(1)\Lambda} & = & - \frac{\Lambda}{9} ( \ x_\mu x_\nu + 2 \eta_{\mu\nu} x^2 \ )
\nonumber \\
h_{\mu\nu}^{(2)\Lambda} & = & \frac{11}{16 \cdot 81} \Lambda^2 ( \ 4 x_\mu x_\nu x^2 + 5 \eta_{\mu\nu} x^4 \ )
\label{323}
\end{eqnarray}

obtained with a very laborious calculation from the Einstein equations.

\section{Conclusion}

The $ AdS $ space-time is an example of curved space-time in the absence of matter. In this paper, we have searched how to describe it with a Lorentz covariant coordinate system, which is better suited for discussing the propagation of gravitational waves in such a curved space.

We first built the perturbative solution of the Einstein equations with a cosmological constant $ \Lambda $ in the Lorentz gauge, and then introduced a more systematic method to construct the solution by embedding the $ AdS $ space-time into a $5d$ space with a quadratic constraint between the coordinates.

We have succeeded to identify the right mapping $ X_a = X_a (x_i) $ which allows us to reach the Lorentz gauge and constructed an iterative method to calculate the $ AdS $ solution to the various perturbative orders in $ \Lambda $. This method is the easiest way to solve the Lorentz gauge condition if compared to the direct perturbative calculation of the Einstein equations.

Once the structure of the $ AdS $ space-time has been accurately identified in a Lorentz covariant coordinate system, we believe that it will be easier to study the propagation of a gravitational wave in the presence of such a background.

\end{document}